\documentclass[useAMS,usenatbib]{mn2e}

\usepackage{amssymb}
\usepackage{amsmath}
\usepackage{graphicx}
\usepackage{textcomp}
\usepackage{array}
\usepackage{multirow}
\usepackage{times}
\usepackage{color}

\def\P{\mathcal{P}}
\def\G{\mathcal{G}}
\def\F{\mathcal{F}}
\def\N{\mathcal{N}}
\def\B{\mathcal{B}}

\def\hm1{h^{-1}}
\def\vpa{{v_\|}}
\def\vx{{\bf x}}
\def\vr{{\bf r}}

\def\vxi{{\bf x_i}}
\def\vs{{\bf s}}
\def\vp{{\bf p}}

\def\xiss{\xi_S(s_\bot , s_\|)\ }

\title[Bivariate Gaussian description for the velocity
PDFs]{Improving the modelling of redshift-space distortions:\\ I. A bivariate Gaussian description for the galaxy pairwise velocity distributions} 
\author[D. Bianchi, M. Chiesa \& L. Guzzo]
{Davide Bianchi$^{1,2}$\thanks{E-mail: davide.bianchi@brera.inaf.it}, Matteo Chiesa$^{1,2}$ \& Luigi Guzzo$^{1}$\\
$^{1}$INAF -- Osservatorio Astronomico di Brera, via Emilio Bianchi
46, I-23807 Merate, Italy\\
$^{2}$Dipartimento di Fisica, Universit\`a degli Studi di Milano, via
Celoria 16, I-20133 Milano, Italy\\
}
\begin{document}

\date{Accepted 2014 October 3. Received 2014 October 3; in original form 2014 July 16}

\volume{446} \pagerange{75--84} \pubyear{2015}

\maketitle

\label{firstpage}

\begin{abstract}
  As a step towards a more accurate modelling of redshift-space distortions
 in galaxy surveys, we develop a general description of the probability distribution function
 of galaxy pairwise velocities within the framework of the
  so-called {\it streaming model}.
For a given galaxy separation $\vr$, such function can be described as a superposition
  of virtually infinite local distributions. We characterize these in terms of
  their moments and then consider the specific case in which they are
  Gaussian functions, each with its own mean $\mu$ and dispersion
  $\sigma$.  Based on physical considerations, we make the further
  crucial assumption that these two parameters are in turn distributed
  according to a bivariate Gaussian, with its own mean and covariance
  matrix. Tests using numerical simulations
  explicitly show that with this compact description one can correctly model
  redshift-space distorsions on all scales, fully capturing the
  overall linear and nonlinear dynamics of the galaxy flow at
  different separations.  In particular, we naturally obtain
  Gaussian/exponential, skewed/unskewed
  distribution functions, depending on separation as observed in
  simulations and data. Also, the recently proposed
  single-Gaussian description of redshift-space distortions is included in this model as a
  limiting case, when the bivariate Gaussian is collapsed to a 
  two-dimensional Dirac delta function.  We also
  show how this description naturally allows for the Taylor expansion of $1+\xi_S(\vs)$ around
  $1+\xi_R(r)$, which leads to the Kaiser linear formula when
  truncated to second order, expliciting its connection with the
  moments of the velocity distribution functions. More work is needed,
  but these results indicate a very promising path to make definitive progress in our
  program to improve RSD estimators.
 
\end{abstract}


\section{Introduction}

Work on the dynamical effect known as ``Redshift Space
  Distortions'' in galaxy surveys \citep[RSD,][]{kaiser1987}, has risen steadily over the
  past few years, following renewed interest in the
  context of the ``dark energy'' problem \citep{guzzo2008,
    zhang2007}.   
Produced by galaxy peculiar velocity flows that are proportional to the growth rate of structure $f$, RSD
provide a potentially powerful way to pinpoint whether a modification of the gravity 
  theory, rather than ``dark energy'', could be  the culprit of the
  apparent acceleration of cosmic expansion. 
 
Estimating $f$ (or, as it is now typical, the combination of the
growth rate
and the amplitude of matter clustering, $f\sigma_8$), however, requires 
cleaning the linear RSD bulk-flow signal from the non-linear components of
the velocity field, modelling their combination
in a sufficiently accurate way.  This is becoming more and more
crucial, given the a-few-percent value of statistical errors already 
reachable by the available largest 
samples \citep[as BOSS,][]{samushia2014, reid2014} and the even
more ambitious expectations for future surveys, as Euclid
\citep{laureijs2011}.  

Until recently, the standard methodology to perform this modelling 
has been based on a modification of the linear theory formalism first
derived in Fourier space by \citet{kaiser1987} and extended to
configuration space by \citet{hamilton1992}, (see \citet{hamilton1998}
for a review).  This ``Dispersion
  Model'', entails convolving the linear model with an
exponential damping term, to account in particular (but not only) for
the evident ``Fingers of God'' small-scale stretching 
due to galaxies in groups and clusters \citep{peacock1999, peacock2001a}.  Despite its simplicity and empirical
basis, this model performs suprisingly well, as shown in a number of
applications
\citep[e.g.][]{peacock2001a,hawkins2003,ross2007,guzzo2008}.  

However,
first accurate tests with simulations showed that the method produces
systematic errors of up to 10\% in the measured values of $f$ ,
depending on the typical mass of the halos in the simulated sample
\citep{okumura2011,bianchi2012}.  These limitations stimulated 
  significant activity to improve RSD modelling
  \citep[e.g.][]{tinker2007, percival2009, white2009, taruya2010, reid2011,
    seljak2011, kwan2012, zhang2013}.   An overview of the different approaches is
  provided in \citet{delatorre2012}.  Several of these developments
 stem from the seminal paper by \citet{scoccimarro2004b}, where (among other
  important developments that we shall encounter later in this paper),
  a more general version of the dispersion model is presented. 
 In such model the linear Kaiser description is
 improved by including contributions from the the galaxy velocity
 divergence power spectrum and the velocity-density cross-power spectrum. 
 A notable development based on the Scoccimarro form is
 represented by the models of  
 \citet{taruya2010} and their implementation in configuration space by
 \citet{delatorre2012}.
 
In the same 2004 paper, Scoccimarro also discusses in general terms the so-called
``Streaming Model'', which in its origins goes back to 
 the early description of peculiar velocities by \citet{davis1983}. 
 \citet{fisher1995} expanded this model into a more general form,
 which was then further generalized in the cited paper by
 Scoccimarro.  This approach has been recently adopted for the
 estimator applied to the
 BOSS DR-9 and DR-11 data releases  \citep{reid2011,reid2012,
   samushia2014}. 

In the present work, we also adopt the streaming
 model as our working framework towards an improved description of RSD.
A particularly appealing feature of this description is that it is
exact, as soon as we have a complete knowledge of the (family of)
Probability Distribution Functions (PDF) of
galaxy pairwise velocities at any galaxy separation in the plane $(r_\bot,
r_\|)$, where $r_\bot$ and $r_\|$ indicate the components of the
separation perpendicular and parallel to the line of sight,
respectively.  Properly describing this family of PDFs is clearly the
  central point in the description of RSD through the
  streaming model.  At a given separation, the corresponding PDF is in fact a pair-weighted average of all
local distributions of galaxy pairs with that separation.  
These local distributions can in principle be completely general.
In practice, they will be governed by the intrinsic properties of the
galaxy flow, which are characterized in general by a bulk
velocity, i.e. a mean streaming component, and a disordered component,
i.e. a dispersion.  This suggests that a sufficiently general description of the 
PDF may be possible, at any $(r_\bot, r_\|)$, in
terms of the first two moments $\mu$ and $\sigma^2$ of the local
distributions.
  
This idea is at the basis of models in which the 
velocity PDF is described by integrating over given functional
forms, e.g. Gaussians. A first example is provided by the work of
 \citet{sheth1996}, who proposes an explanation for the
 nearly-exponential profile 
 of the small-scale pairwise velocity PDF.
This is obtained as a weighted sum of Gaussians, where
 the weighting factor is related to the Press-Schechter multiplicity
 function and to the particle distribution within a clump. 

\citet{tinker2006} and \citet{tinker2007} develop a similar concept in the framework of Halo 
Occupation Models (HOD): redshift-space correlations are described as
the sum of a one-halo and two-halo terms, reflecting respectively the
clustering/virial motions of pairs of galaxies inside halos and the
clustering/relative motions of pairs belonging to different halos.  
The one-halo term is modelled following  a procedure analogue to that of
Sheth, i.e. assuming that satellite galaxy velocities follow a
Gaussian distribution with pairwise dispersion related to the virial
dispersion of the host halo.  The two-halo velocity distribution is a
combination of virial motions inside halos with the halo-halo relative velocities.
The latter  have a distribution
$\P_h$, which is assumed to be described by a superpositions of
Gaussians whose mean and variance are postulated to depend on the 
environment (i.e. local overdensity $\delta$) in which the two halos are found. 
In \citet{juszkiewicz1998}, instead, a skewed
 exponential distribution for the pairwise velocities is constructed
 in the context of Eulerian perturbation theory, based on an
 \textit{ad hoc} ansatz for the pairwise velocity.

In this paper we develop the idea that the two moments of the Gaussian
components behave in fact as jointly
distributed random variables. 
More explicitly, we assume that for any given $(r_\bot, r_\|)$ the
overall velocity distribution can 
be obtained by averaging over a given family of elementary
distributions $\P_L$ (for example, but not necessarily, Gaussian functions)
with statistical weight assigned by the joint probability distribution
$\F(\mu,\sigma)$. 
We shall show that this description is general enough, as to model the
redshift-space correlation function on all scales via the streaming
model. 
Then we focus on the specific case in which $P_L$ and $\F$ are
respectively univariate and bivariate Gaussians, showing that even
under this strong assumption the overall velocity profiles are
correctly reproduced, as well as the corresponding redshift-space
clustering. 
This simple model ultimately shows that the relevant RSD information
is contained into five scale-dependent parameters, namely the mean and
the covariance matrix of the bivariate Gaussian. 
Since the interpretation of these five parameters is clear, they can
be, in principle, expressed as a function of fundamental cosmological
quantities, such as the growth rate of structure.  
We shall explore this connection in a further work.

The paper is organized as follows. 
In Sec. \ref{sec RSD} we introduce our general description of the
line-of-sight pairwise velocity distribution and we discuss its
implications on modelling redshift-space distortions; 
two specific ansatzes for the velocity PDF are discussed in detail:
local Gaussianity and local Guassianity plus global bivariate
Gaussianity;  
in Sec. \ref{sec sims} we test the effectiveness of these assumptions
using N-body simulations; 
our results are summarized in Sec. \ref{sec disco}; 
we discuss perspectives for future developments and applications in
Sec. \ref{sec perspective}.

\section{Dissecting the Streaming Model of Redshift-Space Distortions}\label{sec RSD} 

The streaming model \citep{fisher1995}, in the more
general formulation by \citet{scoccimarro2004b}, describes how the
fractional excess of pairs in redshift space $1 + \xi_S(s_\bot, s_\|)$ is
modified with respect to their real-space counterpart $1 + \xi_R(r)$: 
\begin{equation}
\label{eq streaming}
1 + \xi_S(s_\bot , s_\|) = \int dr_\| \ [1 + \xi_R(r)] \ \P(r_\| - s_\| | \vr) \ .
\end{equation}
Here $r^2=r_\|^2+r_\bot^2$ and $r_\bot=s_\bot$.
This expression is exact: knowing
the form of the pairwise velocity distribution function $\P(\vpa |
\vr) = \P(r_\| - s_\| | \vr)$ at any separation 
$\vr$, a full mapping of real- to redshift-space correlations is
provided.  The knowledge of $\P(\vpa | \vr)$ at any $\vr$ is clearly the key
point in this description. The question to be asked is how general
this knowledge must be, or, in other words, how many degrees of freedom
are necessary for a sufficiently accurate description of this family
of distribution functions and, as a consequence, of RSD.   

The work presented in this paper stems from the attempt to answer this
question. Our purpose is specifically that of finding the minimum set 
of physical parameters, which are still able to predict all the main 
features of the pairwise velocity PDFs along   
the line of sight. This description is required to be accurate enough,
as to recover the correct redshift-space correlation function on all
scales.  The hope is that at a second stage such simplified, yet
accurate, model can be connected to the dynamical description of
RSD and applied to the data to extract information on the growth of
structure (or even broader dynamical quantities characterizing gravity).

\subsection{Characterizing the pairwise velocity distribution
  functions}\label{sec theo}

Let us start with the following general points.  Once a scale $\vr$
is fixed, the distribution function $\P(r_\| - s_\| | \vr)$ that enters
Eq.~(\ref{eq streaming}) could be constructed  -- if we had access to galaxy
velocities -- by building the histogram of the relative velocities of all
pairs with that separation.  If we now imagine to split our Universe
in sub-volumes of appropriate size, by construction we can think without loss of
generality that the overall histogram of pairwise velocities
[i.e. the un-normalized version of $\P(r_\| - s_\| | \vr)$], is given
by the sum of all analogous {\it local}
histograms.  Each of the latter, once normalized,
represents a specific local distribution function
$\P_L(\vpa|\vr,\vxi)$, where $\vxi$ is the location of the
$i$-th sub-volume.  In principle, every $\P_L(\vpa|\vr,\vxi)$ could be
completely different.  In reality, since galaxy dynamics is everywhere
the result of gravitational instability and galaxy velocities in
the different sub-volumes are necessarily correlated, we can reasonably assume that
some fairly general, smooth parametric form is in principle able to
describe the shape of all $\P_L[\vpa | p_1(\vr,\vxi),\dots,p_N(\vr,\vxi)]$, given a set
of $N$ functional parameters $p_j$ to be determined.  

Let us now imagine
the global motions of galaxies within one of the sub-volumes:
physically, it is reasonable to think that on a given scale the 
relative velocities of galaxy pairs can be characterized by the
combination of a
systematic, coherent component (infall onto overdensities or outfflow
from voids) and a random component. In other words, we are
postulating that the local distribution functions can be 
characterized simply by their first two moments, the mean $\mu(\vr,\vxi)$ and
variance $\sigma^2(\vr,\vxi)$. Under these conditions, for any fixed separation $\vr$, we expect the
values of these quantities to be a continuous 
function of the spatial position $\vx$, and therefore described by
their own distribution function (over the sub-volumes).  Let us call it
$\F(\mu,\sigma)$.  Within these assumptions, the global PDF that
enters Eq.~(\ref{eq streaming}), for a given separation
$\vr$, can be expressed as
\begin{equation}
 \label{eq global p}
 \P(\vpa) = \int d\mu d\sigma \ \P_L(\vpa|\mu,\sigma) \ \F(\mu,\sigma) \ .
\end{equation}
The distribution function of $\mu$ and $\sigma$ can be
written as 
\begin{equation}
 \label{eq global F} 
 \F(\mu,\sigma) \equiv \N^{-1} \int d^3x \ A(\vx) \ \delta_D[\mu(\vx)-\mu] \ \delta_D[\sigma(\vx)-\sigma] \ ,
\end{equation}
where $A$ represents the local amplitude, i.e. the local number density of pairs\footnote{
For any given separation $\vr$, we can define the number density of pairs as
$A(\vx) = \left[1+\delta\left(\vx-\frac{\vr}{2}\right)\right] \left[1+\delta\left(\vx+\frac{\vr}{2}\right)\right]$
where $\delta$ is the number-density contrast.
We then obtain $\N=1+\xi(r)$.}, $\N = \int d^3x \ A(\vx)$ and
$\delta_D$ are Dirac delta functions.
By substituting Eq.~(\ref{eq global F}) into Eq.~(\ref{eq global p}), we obtain
\begin{equation}
 \label{eq global P w} 
 \begin{split}
  \P(\vpa) & = \N^{-1} \int d^3x d\mu d\sigma \ \P_L(\vpa|\mu,\sigma) \\
  & \times \ A(\vx) \ \delta_D[\mu(\vx)-\mu] \ \delta_D[\sigma(\vx)-\sigma] \\
  & = \N^{-1} \int d^3x \ A(\vx) \ \P_L[\vpa|\mu(\vx),\sigma(\vx)] \ ,
\end{split} \end{equation} which clarifies that we are in fact
modelling the global PDF as a pair-weighted mean of a (normalized)
functional form 
parameterized by its first two moments. 
We define the mean of $\mu$ and $\sigma$
in a compact form, \begin{equation} 
 M_k \equiv \int d\mu d\sigma \ \mu^{1-k} \sigma^k \ \F(\mu,\sigma) \ ,
\end{equation}
where $k \in \{0,1\}$, i.e. $M_0$ and $M_1$ represent the mean of $\mu$ and $\sigma$, respectively.
Similarly, we define the (tensorial) central moments,
\begin{equation}
 \begin{split}
  C_{k_1,\cdots,k_n}^{(n)} \equiv \int d\mu d\sigma \ {(\mu-M_0)}^{n-\sum_i k_i} \times & \\ \times \ {(\sigma-M_1)}^{\sum_i k_i} \ \F(\mu,\sigma)
 \end{split}
\end{equation}
where $k_i \in \{0,1\}$ and $n=0,1,2,\ldots$ is the order of the tensor.
Trivially, $C^{(0)}=1$ and $C^{(1)}=(0,0)$.
We shall denote the moments and central moments of order $n$ of $\P$ as $m^{(n)}$ and $c^{(n)}$, respectively.
Finally, we shall adopt the same notation, but adding a subscript $L$,
to describe the moments of the $\P_L$.  To ease comprehension in the
development of the paper, all definitions are summarized in compact
form in Table \ref{tab def}.
\begin{table*}
 \centering
  \begin{tabular}{cll}
  &&\\
   \multicolumn{1}{c}{\Large PDF} & \multicolumn{1}{c}{\Large moments} & \multicolumn{1}{c}{\Large central moments} \\
  \hline
  \hline
  &&\\
  $\P$ & $m^{(n)} \equiv \int d\vpa \ \vpa^n \ \P(\vpa)$ & $c^{(n)} \equiv \int d\vpa \ \left(\vpa-m^{(1)} \right)^n \ \P(\vpa)$ \\
  &&\\
  $\P_L$ & $m^{(n)}_L \equiv \int d\vpa \ \vpa^n \ \P_L(\vpa)$ & $c^{(n)}_L \equiv \int d\vpa \ \left(\vpa-m^{(1)}_L \right)^n \ \P_L(\vpa)$ \\
  &&\\
   $\F$ & $M_k \equiv \int d\mu d\sigma \ \mu^{1-k} \sigma^k \ \F(\mu,\sigma)$ & $C_{k_1,\cdots,k_n}^{(n)} \equiv \int d\mu d\sigma \ {(\mu-M_0)}^{n-\sum_i k_i} {(\sigma-M_1)}^{\sum_i k_i} \ \F(\mu,\sigma)$ \\
  &&\\
 \end{tabular}
 \caption{Definitions and notation adopted to describe the moments of
   the three probability distribution functions (PDFs) considered in
   this work: $\P$, $\P_L$ and $\F$. 
   $n$ is the order of the moment and $k \in \{0,1\}$. 
 Throughout the text $\mu=m^{(1)}_L$ and $\sigma^2=c^{(2)}_L$.
 Since we do not need to define $n$-th order (non-central) moments of
 $\F$, it is intended that $M_k=M_k^{(1)}$.}
 \label{tab def}
\end{table*}
From Eq. (\ref{eq global p}), it follows that
\begin{equation}
 m^{(n)} = \langle m^{(n)}_L \rangle \ ,
\end{equation}
where $\langle \ \dots \rangle \equiv \int d\mu d\sigma \dots \ \F(\mu,\sigma)$.
On the other hand
\begin{equation}
 c^{(n)} \neq \langle c^{(n)}_L \rangle \ ,
\end{equation}
which implies that, for example, it is possible to obtain a skewed
global distribution $\P$ by superposition of non-skewed local
distributions $\P_L$, as we show in more detail in the following. 

\subsection{The velocity distribution functions as a
  Gaussian family of Gaussian functions 
  }\label{sec GauGau} 

Let us then derive first the expressions of the first few moments of
the global distribution $\P$ under completely general conditions\footnote{Note that in
  the literature, $m^{(1)}$ and $c^{(2)}$ have often been denoted with
  $v_{12}$ and ${\sigma_{12}}^2$, respectively, see \citet{fisher1994b}.
}. The full derivation is presented in Appendix~\ref{app moms}. The
results are reported in the upper part of Table~\ref{tab moms}. 
\begin{table*}
 \centering
 \begin{tabular}{clll}
  &&&\\
  & \multicolumn{1}{c}{\Large $n$} & \multicolumn{1}{c}{\Large $m^{(n)}$} & \multicolumn{1}{c}{\Large $c^{(n)}$} \\
  \hline
  \hline
  \parbox[t]{2mm}{\multirow{10}{*}{\rotatebox[origin=c]{90}{\large general}}}
  &&&\\
  & 0 & $1$ & $1$ \\
  &&&\\
  & 1 & $M_0$ & $0$ \\
  &&&\\
  & 2 & ${M_0}^2 + {M_1}^2 + C^{(2)}_{00} + C^{(2)}_{11}$ & ${M_1}^2 + C^{(2)}_{00} + C^{(2)}_{11}$ \\
  &&&\\
  & 3 & ${M_0}^3 + 6 M_1 C^{(2)}_{01} + 3 M_0 \left({M_1}^2 + C^{(2)}_{00} + C^{(2)}_{11}\right)+ \qquad$ & $6 M_1 C^{(2)}_{01} + C^{(3)}_{000} + 3C^{(3)}_{011} + \langle c^{(3)}_L \rangle$ \\
  &   & $\quad +C^{(3)}_{000} + 3C^{(3)}_{011} + \langle c^{(3)}_L \rangle$ & \\
  &&&\\
  \hline
  \parbox[t]{2mm}{\multirow{15}{*}{\rotatebox[origin=c]{90}{\large GG}}}
  &&&\\
  & 0 & $1$ & $1$  \\
  &&&\\
  & 1 & $M_0$ & $0$ \\
  &&&\\
  & 2 & ${M_0}^2 + {M_1}^2 + C^{(2)}_{00} + C^{(2)}_{11}$ & ${M_1}^2 + C^{(2)}_{00} + C^{(2)}_{11}$ \\
  &&&\\
  & 3 & ${M_0}^3 + 6 M_1 C^{(2)}_{01} + 3 M_0 \left({M_1}^2 + C^{(2)}_{00} + C^{(2)}_{11}\right)$ & $6 M_1 C^{(2)}_{01}$ \\
  &&&\\
  & 4 & ... & $3 \left({M_1}^2 + C^{(2)}_{00} \right)^2 + 6 \left[C^{(2)}_{11} \left(3{M_1}^2 + C^{(2)}_{00}\right)+ 2{C^{(2)}_{01}}^2\right] + 9 {C^{(2)}_{11}}^2$ \\
  &&&\\
  & 5 & ... & $60 M_1 C^{(2)}_{01} \left({M_1}^2 + C^{(2)}_{00} + 3
    C^{(2)}_{11}\right)$ \\
  &&&\\
 \end{tabular}
 \caption{Expressions for the moments of the velocity distribution
   $\P(\vpa)$ as a function of the moments of $\F$, in the most
   general case (upper panel) and under the stronger GG assumption discussed in
   the text. In the latter case, we also report the 4-th and 5-th
   central moment since the set of equations with $2 \le n \le 5$ can
   be inverted to recover $C$ and $M$ as a function of $c$ and $m$.} 
 \label{tab moms}
\end{table*}
A very important outcome to be noted from these computations is that
even if the skewness of the local $\P_L(\vpa|\mu,\sigma)$ and of the
distribution $\F$ are negligible (i.e. we set $c^{(3)}_L=0$ and 
$C^{(3)}_{000}=3C^{(3)}_{011}=0$, respectively), we can still obtain a global
skewness, $c^{(3)}=6 M_1 C^{(2)}_{01}$, as the result of  the (pair-weighted)
covariance between the two moments $\mu$ and $\sigma$.  This is a remarkable result,
as it suggests that a simple, symmetric (i.e. unskewed) shape for $P_L$
and $\F$ would be sufficient to describe without a large loss of
generality the overall pairwise velocity distribution $\P$.  

Let us therefore assume a Gaussian form for the local distribution functions
$\P_L$, i.e. 
\begin{equation}
 \P_L = \G(\vpa|\mu,\sigma) = \frac{1}{\sqrt{2\pi}\sigma}
 \exp\left[{-\frac{{(\vpa-\mu)}^2}{2 \sigma^2}}\right] \ ,
\end{equation}
 such that the overall $\P(\vpa) $ is written as 
\begin{equation}
\label{eq gau1}
 \P(\vpa) = \int d\mu d\sigma \ \G(\vpa|\mu,\sigma) \ \F(\mu,\sigma) \ .
\end{equation}
We shall refer to this as the ``local Gaussianity'' (LG)
assumption. 

We then assume that the bivariate
distribution of the $\mu$ and $\sigma $ parameters describing these
Gaussians is in turn a bivariate Gaussian. This corresponds to saying
that the pair-weighted distribution $\F$ is given by
\begin{equation}
\label{eq gau1biv2}
 \P(\vpa) = \int d\mu d\sigma \ \G(\vpa|\mu,\sigma) \ \B(\mu,\sigma) 
\end{equation}
where
\begin{equation}
\B(\mu,\sigma) = \dfrac{1}{2 \pi \sqrt{\det(C)}} \exp
\left[-\frac{1}{2} \Delta^T C^{-1} \Delta \right]
\end{equation}
and
\begin{equation}
 \Delta =
 \begin{pmatrix}
  \mu - M_0 \\
  \sigma - M_1
 \end{pmatrix}
 \qquad
 C =
 \begin{pmatrix}
  C^{(2)}_{00} & C^{(2)}_{10} \\
  C^{(2)}_{01} & C^{(2)}_{11} 
 \end{pmatrix} \ ,
\end{equation}
with $C^{(2)}_{10} = C^{(2)}_{01}$.
We shall refer to the this second assumption (Eq. (\ref{eq gau1biv2}))
as ``Gaussian (local) Gaussianity'' (GG). In the following
section we shall test directly the validity of both LG and GG assumptions.

In the lower part of Table \ref{tab moms} we report the expressions
that are obtained for the first few moments of $\P$ under the GG
assumptions, as discussed in the Appendix.  Also on these aspects
there is ample room for further developments that are not explored
here.  In a work in preparation we are investigating a theoretical
prescription for the dependence of $M$ and $C^{(2)}$ on $\vr$; in
this framework it can also be shown that all moments can be computed up
to any order through a moment generating function.

\begin{figure*}
 \begin{center}
   \includegraphics[width=14cm]{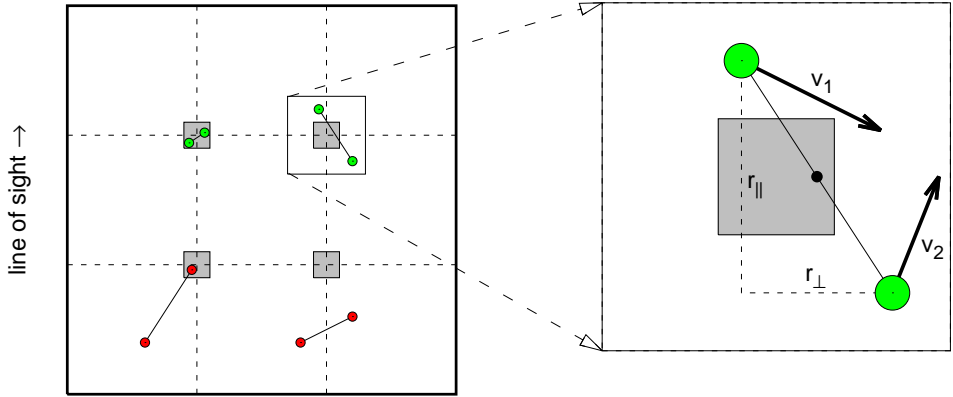}
    \caption{Two dimensional sketch of the procedure adopted to measure the local moments $\mu$ and $\sigma^2$ (i.e. $m_L^{(1)}$ and $c_L^{(2)}$) from the simulation.
    Gray filled squares represent ${(10\hm1\text{Mpc})}^3$ sub-volumes, ideally corresponding to different local realizations (the scale and the number of sub-volumes in the figure are arbitrary).
    For each local realization, only pairs with mid-point falling inside the corresponding gray area are included in the calculations (green).
    Pairs not satisfying this requirement are excluded from the statistics (red). 
    For each green-like pair we measure separation $(r_\bot,r_\|)$ and line-of-sight pairwise velocity $\vpa$.
    We then use these informations to build different local estimations (one per each sub-volume) of the velocity distribution (and its moments) as a function of the separation.}
  \label{fig grid}
 \end{center}
\end{figure*}

\section{Testing the accuracy of the LG and GG assumptions}\label{sec sims}
To test the goodness of the LG and GG descriptions developed in the
previous section, we use a properly
chosen numerical simulation to directly measure the distribution of
pairwise velocities at different separations.
It is important to note that in this exercise we are not just checking whether the
functional form of Eq. (\ref{eq gau1biv2}) is general enough to
describe $\P$, for any given $\vr$, by fitting for the mean $M$ and
the covariance $C^{(2)}$ of $\B$ as free parameters. 
Rather, we want to make sure that these quantities have a well
defined physical interpretation by directly measuring $\mu$ and
$\sigma$ from galaxy velocities in the simulation. 
If so, a full theoretical prediction for $\B$ is in principle
feasible.

For our test we use the data from the MultiDark Bolshoi run \citep{riebe2013}, at $z=0$.
Assuming a set of comological parameters compatible with WMAP5 and
WMAP7 data, $\{\Omega_m, \Omega_{\Lambda}, \Omega_b, \sigma_8, n_s\} =
\{0.27, 0.73, 0.047, 0.82, 0.95\}$, this N-body simulation follows the
dynamics of $2048^3$ particles over a cubical volume of ${(250 h^{-1}
  \text{Mpc})}^3$. 
If the goal were to test how accurately a given RSD model can recover the
underlying cosmology (e.g. the growth rate of structure), such volume
would probably be too small. 
For our scope, however, a small, high-resolution simulation is the
best choice, as we are interested in testing in detail how the LG and
GG models are capable to recover the ``true'' overall velocity PDF and
redshift-space correlation function, given the measured local
PDFs.  

\subsection{Practical estimate of the local velocity distribution
functions}
The strategy adopted to measure the local distribution $\P_L$ is sketched in Fig. \ref{fig grid}.
We consider a grid with $N_L=11^3$ nodes, which ideally correspond to
$N_L$ local realizations, i.e. the sub-volumes discussed in
Sec. \ref{sec theo}. 
$N_L$ is basically limited by the amount of available RAM. 
Since the CPU time depends essentially on the number of particles, we
randomly dilute the sample down to $\approx 1.4 \times 10^7$
particles.
We then store $v_\|$ for all pairs whose mid-point falls inside a $10
\hm1 \text{Mpc}$ cube centered on the given $i=1,2,\dots,11^3$ grid
node (see Fig. \ref{fig grid})\footnote{The continuous limit is readily
  obtained by considering a denser grid (i.e. larger $N_L$) with nodes
  surrounded by smaller cubes.}. 
Given a separation $(r_\bot,r_\|)$, we then compute $\mu_i$, $\sigma_i$ and 
$A_i$, assuming the plane parallel approximation and rotational
symmetry around the line of sight. 
We adopt $1 h^{-1} \text{Mpc}$ bins for both separation $(r_\bot,r_\|)$ and velocity $v_\|$.
In order to avoid discretization effects, for any given grid node (sub-volume) and
separation, the corresponding $\P_L$ is included in the calculations
only if sampled by more than $100$ pairs.
We have checked that the results reported in the following do not
depend on this particular choice by repeating the procedure for
different pair thresholds. 
From a theoretical point of view, it seems clear that any possible
dependence on the threshold is mitigated by the fact that $\F$ is a
pair-weighted distribution. 
This guarantees that poorly sampled local distributions do not
contribute much to the global description. 

\subsection{Results}
\begin{figure*}
 \begin{center}
   \includegraphics[width=16cm]{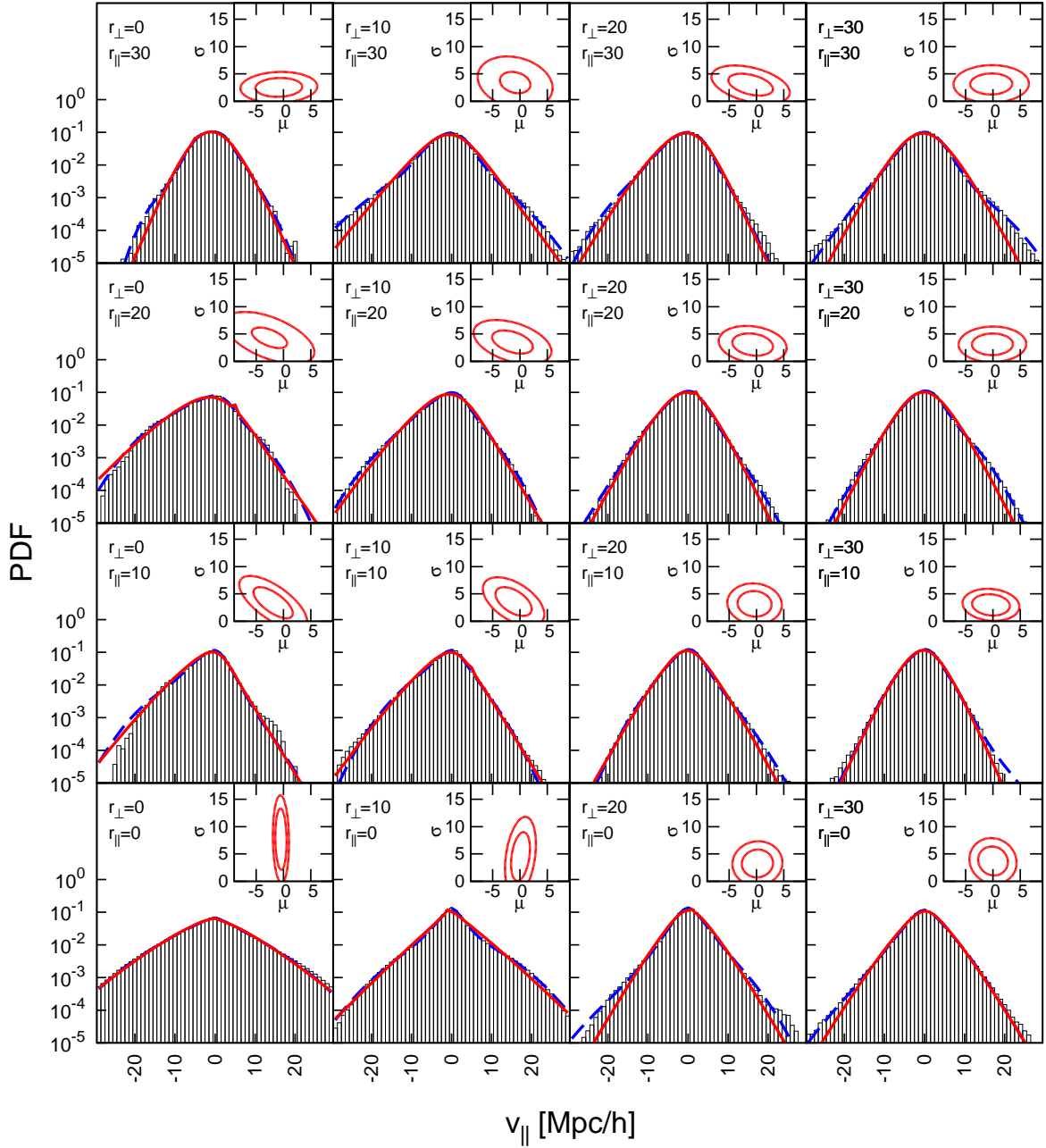}
    \caption{The distribution functions of line-of-sight pairwise velocities 
      $\P(\vpa)$, measured from the $z=0$ snapshot of the simulation at a few selected
      separations as described in the text; 
      the values of the separation $(r_\bot, r_\|)$ are given in each
      panel in units of $h^{-1}\text{Mpc}$.  The superimposed dashed
      and solid lines
      give the model PDFs for the LG and GG cases, i.e.: (a) assuming that at
      any separation the local distributions $\P_L$ are
      described by Gaussian functions, for which the two moments
      $\mu_i$ and $\sigma_i^2$ are masured and used to empirically build the
distribution function $\F$ (LG, blue dashed line); (b) making the further
assumption that $\F$ is described by a bivariate Gaussian
$\B(\mu,\sigma)$ as given by eq.~\ref{eq gau1biv2} (GG, red solid line).
The $1\sigma$ and $2\sigma$ contours of the
$\B(\mu,\sigma)$  corresponding to each separation $(r_\bot, r_\|)$ are shown in the
upper right corner of each panel. 
Also for $\mu$ and $\sigma$ units are $h^{-1}\text{Mpc}$.
    } 
  \label{fig PDF_superposed}
 \end{center}
\end{figure*}

In Fig. \ref{fig PDF_superposed} we plot the results of our test for 16 representative
values of the pair separation $(r_\bot,r_\|)$ within the range $[0,30]$
$h^{-1}$ Mpc. Each panel in the figure shows the measured histogram describing the 
pairwise velocity distribution $\P(\vpa)$ for pairs at that separation, comparing
it with the two models of the same quantity constructed under the LG (blue dashed curve)
and GG (red solid curve) assumptions.  The 
panels provide an interesting overview of how the morphology of the
velocity PDF can
vary depending on the relative separation and orientation of the
galaxy pairs, changing from exponential to Gaussian, with or without
skewness.  As we immediately notice from the dashed and solid lines,
however, this apparent complexity is in general captured
quite well also under the simplifying assumptions of our models. 

The agreement between the
data histogram and the blue dashed line describing the LG model is
indeed excellent at all separations and over the full range of pairwise velocities. This
validates our first assumption, i.e. that the local 
distribution functions that concur to form the global one at a given
separation are well described as a family of Gaussians.  
Also in the case of the red solid curves, corresponding to the
stronger assumption that the two parameters $\mu$ an $\sigma$ describing
these Gaussians are also Gaussian distributed, the
model PDF follows the data histogram very well. Some discrepancy
is visible only at the tails of the distributions and for some
separations, and is discussed below.  We note that to build the PDFs
under the LG assumption we  have
combined (i.e. summed and then normalized) all Gaussians defined by the
values of (local) mean $\mu_i$, variance $\sigma_i^2$ and
amplitude $A_i$ measured at each grid node.  This is preferable to
estimating the distribution of the moments $\F$ by constructing 
a two dimensional histogram of $\mu$ and $\sigma$, as Eq. (\ref{eq
  gau1}) would require. 

Conversely, to test the GG assumption we actually estimate $\B$ (i.e. its mean
and covariance matrix) from the simulation and the overall
distribution $\P(\vpa)$ is then obtained by simply applying Eq.(\ref{eq
  gau1biv2}). 
The numerical estimate of $\B$ inevitably adds some instability to the
resulting PDFs in the GG case, which is probably at the origin of the
small discrepancies observed with respect to the LG curves.  
The $1-$ and $2-\sigma$ contours of $\B(\mu,\sigma)$, which univocally
determine $\P(\vpa)$ under the GG assumption, are shown as insets in
each panel of Fig. \ref{fig PDF_superposed}.  We note
that the majority of the $\B$ distribution is in fact contained inside the $\sigma>0$
plane. This is a crucial validation of the model, given that the
region $\sigma \le 0$ corresponds to an unphysical negative velocity 
dispersion.  

The position and orientation of the
elliptical contours of $\B(\mu,\sigma)$ vary in each panel and have important
physical implications, in connection to Table \ref{tab moms} (lower right
panel).  
The $\mu$ value of the centre of the ellipse corresponds to the mean
of the distribution $\P(\vpa)$.
On the other hand, the variance of $\P$ is related to the combination of the
$\sigma$ value of the centre of the ellipse with, roughly speaking,
its size (as described in the third line of the GG panel of Table~\ref{tab moms}). 
Finally, the skewness of $\P(\vpa)$ is due to the covariance between $\mu$
and $\sigma$, which corresponds to the rotation of the ellipse with
respect to the Cartesian axes.
For separations with $r_\|=0$ $h^{-1}$ Mpc,
the axes of the ellipses are subtsantially aligned with the Cartesian
axes, and are centred at $\mu=0$. This is expected from isotropy
considerations and corresponds to zero skewness and mean of the pairwise
distributions, as evident from the histograms.  In particular, for the
bin at separations $(0,0)$ the ellipse is very narrow, with very little
variance on $\mu$ and large variance on $\sigma$. This is what one
expects if the distribution at these separations is essentially
dominated by virialized regions, for which the mean streaming is
negligible.  

\begin{figure*}
 \begin{center}
   \includegraphics[width=14cm]{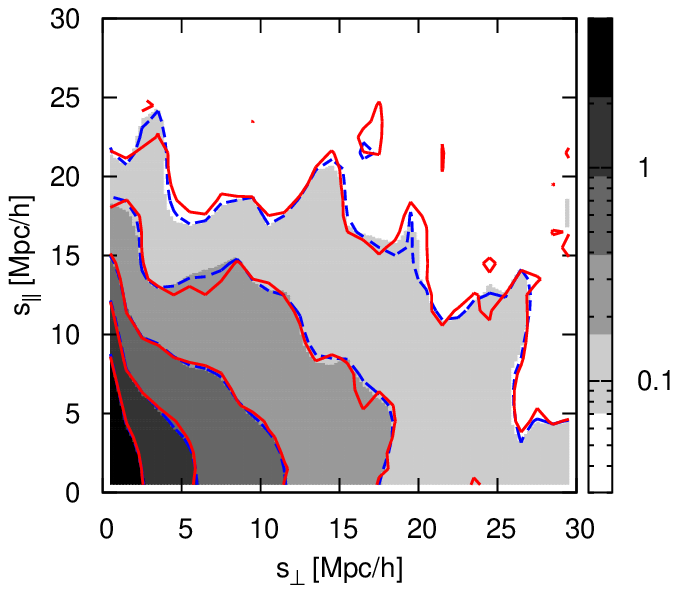}
    \caption{The redshift-space correlation function
      $\xiss$ 
measured at $z=0$ from the simulated sample as described in the text.  The
grayscale contours correspond to the direct measurement; the blue
dashed contours correspond to fitting each local distribution of
pairwise velocities $\P_L$ with a Gaussian function and measuring its
two moments $\mu_i$ and $\sigma_i^2$ to empirically build their
distribution function $\F$; the red solid curves are instead based on the
further assumption that $\F$ is described by bivariate Gaussian, 
as described by $\B(\mu,\sigma) $ in eq.~\ref{eq gau1biv2}. In
practice, the contours demonstrate the impact of reducing the degrees
of freedom in the form of the distribution function of pairwise
velocities. The level of fidelity of the red solid contours when
compared to the gray-scale ones shows the goodness of the bivariate
Gaussian assumption. Note that the ``unsmoothed'' appearance of $\xiss$
is not at all an issue, reflecting the limited
number of ``local samples'' involved in the specific evaluation. The
goal of this exercise is to show that the same $\xiss$ can be
obtained when using the directly measured velocity distribution, or
its modelization under the increasing assumptions of the LG and GG models  (see text). 
}
  \label{fig xis}
 \end{center}
\end{figure*}
Finally, in Fig. \ref{fig xis}  we test how these assumptions for the
velocity distribution perform when modelling redshift-space
distortions through the streaming model
(Eq. \ref{eq streaming}). In this figure, we
compare the redshift-space correlation function $\xiss$ from the
simulation (gray scale) to those obtained with the LG (blue dashed
contours) and GG (red solid contours) models.  In practice, given the real-space 
correlation function $\xi_R$, we use the
streaming model to compute the $\xiss$
corresponding to the three different  distributions $\P$ previously
discussed (direct measurement, LG and GG).  To minimize discretization
effects, we use a smooth 
$\xi_R$ obtained by Fourier transforming the power spectrum given by
CAMB \citep{lewis2000} for the same cosmology of the Bolshoi simulation. 
Despite our PDFs are too poorly sampled to yield a smooth $\xi_S$, it
is evident that the direct measurement is very accurately described
by the LG approximation (blue dashed contours). 
As in the previous figure, the GG model (red solid contours), seems to
give a slightly less stable $\xi_S$, nonetheless the agreement with
the direct estimate remains very convincing. 
Note that the ``unsmoothed'' appearance of $\xiss$
is no cause for concern, as it simply reflects the limited
number of ``local samples'' involved in the specific
evaluation. A larger box together with a higher sampling over a denser grid (i.e. larger $N_L$) would
produce a smoother function at the cost of 
a much heavier computational effort. However, this has no
real justification for our purpose as the key point here is to test how well the
dashed and solid contours reproduce the directly measured $\xiss$ {\it
  for the same sample},
including its details and deviations around the general expected form. This
goal is clearly achieved here.
 
\section{Connection of the streaming model with the moments of $\P$}\label{sec Streaming} 

Important insight on the role played by the moments of the
overall distribution $\P$ in the description of redshift-space
clustering, can be gained by going back to the expression of the
streaming model.  

 By substituting Eq.(\ref{eq global p}) in Eq.(\ref{eq streaming}) and
 making the dependence of $\F$ on $\vr$ explicit, we obtain 
\begin{align}
\label{eq streamingF}
& 1 + \xi_S(s_\bot , s_\|) = \nonumber \\
= & \int d\mu d\sigma \int dr_\|\ [1 + \xi_R(r)] \ \P_L(r_\| - s_\| |\mu,\sigma) \ \F(\mu,\sigma|\vr) \ .
\end{align}
Let us then Taylor expand the term $(1+\xi_R) \times \F$ around $r_\|=s_\|$:
\begin{align}
& 1 + \xi_S(s_\bot , s_\|) = \nonumber \\
= & \sum_n\frac{1}{n!} \int d\mu d\sigma \ \int  dr_\| \ {(r_\| -
  s_\|)}^n \P_L(r_\| - s_\| |\mu,\sigma) \nonumber \\ 
& \times \frac{\partial^n}{\partial {r_\|}^n} \{ [1 + \xi_R(r)]
\F(\mu,\sigma|\vr) \} \Big|_{r_\|=s_\|} \nonumber
\end{align}
\begin{align}
\label{eq expansion}
= & \sum_n\frac{1}{n!} \int d\mu d\sigma \ m_L^{(n)}(\mu,\sigma)
\frac{\partial^n}{\partial {r_\|}^n} \{ [1 + \xi_R(r)]
\F(\mu,\sigma|\vr) \} \Big|_{r_\|=s_\|} \nonumber \\ 
= & \sum_n \frac{1}{n!} \frac{\partial^n}{\partial {r_\|}^n} \left\{
  [1 + \xi_R(r)] \langle m_L^{(n)} \rangle \right\} \Big|_{r_\|=s_\|}
\nonumber \\ 
= & \sum_n \frac{1}{n!} \frac{\partial^n}{\partial {r_\|}^n} \left\{
  [1 + \xi_R(r)] m^{(n)}(\vr) \right\} \Big|_{r_\|=s_\|} \ . 
\end{align}
Interestingly, this is the same expansion derived by \citet[][Eq. 53]{scoccimarro2004b}, following a slightly different procedure.

There are a few important aspects to underline. First, we note that
expression~(\ref{eq expansion})
does not depend on the number of moments we consider, i.e. if 
$\P_L=\P_L(\vpa|m_L^{(1)},c_L^{(2)},\dots,c_L^{(k)})$ and
$\F=\F(m_L^{(1)},c_L^{(2)},\dots,c_L^{(k)}|\vr)$, Eq. (\ref{eq
  expansion}) still holds for any $k$. 
Furthermore, the generic term of order $n$ depends only on the first
$n$ local moments\footnote{This means that if, for example, $k=2$, the
  first two term of the expansion do not depend on the particular
  functional form chosen for $\P_L$.}. 
More in general, 
this expression holds if the system can be
statistically modeled in terms of a set of distributions parameterized by one
or more random variables.  In our specific development of the GG model
in section~\ref{sec GauGau} we
have considered the case in which the random variables 
are $\mu=m_L^{(1)}$ and $\sigma=\sqrt{m_L^{(2)}}$.
Eq. (\ref{eq expansion}) can be read as a natural 
expansion of the redshift-space correlation function around the
real-space correlation function, which corresponds to the $n=0$
term.

Finally, if in Eq. (\ref{eq expansion}) we limit the expansion to $n=2$ and assume
that $1+\xi_R \approx 1$ and $\partial^n \xi_R / \partial r^n_\|
\approx 0$ (both reasonable assumptions at large
separations -- but see below), we obtain
\begin{equation}
 \label{eq Kaiser limit}
 \xi_S(s_\bot,s_\|) = \xi_R(s) + \frac{\partial}{\partial s_\|}m^{(1)}(\vs) + \frac{1}{2} \frac{\partial^2}{\partial s^2_\|}m^{(2)}(\vs) \ .
\end{equation}
As shown by \citet{fisher1995} and \citet{scoccimarro2004b}, this expression corresponds to Kaiser's linear model
\citep{kaiser1984}.
Eq. (\ref{eq expansion}), therefore, 
naturally includes the Kaiser linear limit as a specific case.
It is interesting to note that the condition that $\partial^n \xi_R / \partial r^n_\|
\approx 0$ implies that, despite being on scales $\sim 110$ h$^{-1}$
Mpc, it might be problematic to apply  
the Kaiser limit around the BAO peak, since there the derivative of
$\xi_R$ is far from being zero.

This derivation suggests a further interesting way to make progress in
the modelling of RSD, with respect to the one 
developed in section~\ref{sec GauGau}.  This would entail extending the proposed
streaming-model expansion of Eq. (\ref{eq expansion}) to
some arbitrary order larger than the Kaiser ($n=2$) limit, until the
description is satisfactory.\footnote{Such approach
  requires to ensure that the expansion is perturbative,
  i.e. it exists a range of separations over which the $(n+1)$-th term
  is smaller than the $n$-th. This should be verified theoretically
  and against simulations.
}  
 We plan to explore this approach in a future work.

\section{Summary and Discussion}\label{sec disco}

Based on quite general statistical considerations, we have developed a
simple model for the galaxy pairwise velocities distribution along the
line of sight $\P(\vpa)$, in which, at each separation, $\P$ is
written as the pair-weighted mean of local distributions
$\P_L(\vpa|\vp)$ (where $\vp$ represents an arbitrary set of
parameters).  
More explicitly,  $\P(\vpa) = \int d\vp \ \P_L(\vpa|\vp) \ \F(\vp)$,
where $\F(\vp)$ is the overall pair-weighted joint distribution
function of the parameters $\vp$. 

A general relation between the moments of $\P$ and $\F$, is provided
for the specific case in which the parameters are the pairwise
velocity mean $\mu$ and standard deviation $\sigma$.  We have shown
that the ``true'' overall velocity distribution $\P$ is recovered on
all scales under the simple assumption that the local distributions
$\P_L$ are Gaussian functions whose mean $\mu$ and standard deviation
$\sigma$ are in turn distributed according to a bivariate Gaussian.
This corresponds in practice to compressing the whole RSD information
into five well-defined physical parameters, i.e. the two central
values plus three independent elements of the covariance matrix of the
bivariate Gaussian.  This can be seen as a natural extension to the
recently proposed pure Gaussian descriptions of RSD \citep{reid2011},
which can be obtained from our model as the limiting case in which the
bivariate expression becomes a two-dimensional Dirac delta. 

Our approach allows the redshift-space
correlation function to be expanded in terms of the individuals moments of the
distribution $\P$, independently of the shape of $\P_L$ and
$\F$, expliciting the contribution of such moments to the description of
redshift-space clustering. 
We have seen how this expansion recovers the 
Kaiser limit at large separations.  
Both approaches, bivariate Gaussian description and streaming-model
expansion, suggest possibile developments in the modelling of RSD, which we
discuss below in Sec. \ref{sec perspective}.  

We have seen in the introduction that the idea of describing  the
velocity PDFs by integrating over given functional forms,
e.g. Gaussians, is not new in the literature.  The approach presented
here has similarities to previous works, but also important
differences.  The main novelty is in the bivariate Gaussian
form, which we have shown is capable to reproduce quite well the measured
velocity PDF at all separations. This includes a ``natural'' skewness,
which is encapsulated by the covariance between the mean and variance of
the family of Gaussians.  More specifically, \citet{sheth1996} obtains the nearly-exponential profile 
 of the small-scale pairwise velocity PDF by adding Gaussians, which
 are weighted by a factor related to the Press-Schechter multiplicity
 function and to the particle distribution within a clump.  This  is
 formally quite different from the description developed 
 here. It is also valid for highly non-linear scales only,
 whereas our approach is fully general.  

A comparison with the work of Tinker and collaborators
\citep{tinker2006,tinker2007} is not straightforward. However, our
model for $\P$ could mimic their prescription for the distribution
function of halo-halo relative
velocities, $\P_h$, e.g. if we assumed that the dependence of
$\mu$ and $\sigma$ on position 
is totally driven by the local value of the density field.  Very
recently, a similar HOD approach combined with simulations has been
applied to the 
SDSS-DR10 data, pushing the analysis to very small scales and obtaining a
very precise estimate of the growth rate product 
$f\sigma_8$ \citep{reid2014}. 

In the work of \citet{juszkiewicz1998}, instead, the skewness of the
PDF is shown to arise as a consequence of the non-trivial 
 cross-correlation between velocity and density and 
 some similarity with our approach is encoded in their Eq. (13), where
 the pairwise velocity distribution  is described as a weighted sum
 of Gaussians.  Such equation is a direct consequence of a specific
 choice of the PDF of the density
 contrast.  Our approach, however, is based on completely 
 different assumptions, as here we do not make (yet) any hypothesis on the
 density field nor follow any perturbative scheme.

Finally, the work of \citet{zu2013}, which models the cluster-galaxy
cross-correlation function in redshift space, presents some possibly interesting
connections to our approach.  In that paper,  
given a cluster-galaxy separation $r_{cg}$, the
joint distribution $\P_{2D}(v_r,v_t|r_{cg})$ is considered. Here $v_r$ and $v_t$
are the components of the pairwise velocity parallel and
perpendicular to the separation vector $\vr_{cg}$, respectively. 
$\P_{2D}(v_r,v_t|r_{cg})$ is then described as the combination of a
virialized term with an isotropic Gaussian velocity distribution and
an infall term modelled as a skewed 2D $t$-distribution. 
The line-of-sight distribution $\P(v_{||}|\vr_{cg})$ is then obtained
from $\P_{2D}(v_r,v_t|r_{cg})$ by projection. 
A direct comparison with our model is not feasible, since we are
dealing with auto-correlations rather than cross-correlations; still, we could
imagine of adopting a similar two-dimensional approach, e.g. by modeling
$\P_{2D}$ as the product of two components: a full 5-parameter
GG-distribution for $v_r$ and a simplified 3-parameter GG-distribution
for $v_t$ (in the tangential direction $M_0$ and $C^{(2)}_{01}$ vanish
due to isotropy). 
In practice, with this approach the angular dependence could be
naturally removed at the cost of adding three more parameters, which
could be an interesting test.

\subsection{Perspectives}\label{sec perspective}

Our description of the velocity PDF opens a number of interesting
questions. The definition of a full model of RSD that
can be applied in practice to observational data obviously requires
further developments.  These include in particular building the connection to
the dynamics of clustering and thus to the growth rate of structure
and related cosmological parameters.  
While we already started working on this part of the modelling and
expect results to be presented in a future paper, it is useful here to
sketch some general considerations on different ways through which we
expect this to be feasible. 

A model-dependent possibility could be to use simulations to estimate how our
 bivariate Gaussian description deviates from the two-dimensional
 Dirac delta corresponding to
 the simple Gaussian model of \citet{reid2011}. This would provide in
 practice an empirical nonlinear correction to such model. 
 This can be seen a sort of configuration-space extension of the
 Fourier-space approach proposed by \citet{kwan2012} and
 \citet{vallinotto2013}. However, the physical meaning of the parameters in
 our interpretation would be completely different. 

Another possibility would be to provide a theoretical prediction for the
  bivariate Gaussian, i.e. to derive the equations for the
  five parameters on which it depends (the mean values and its
  covariance matrix). This is in principle feasible
  and some analogy might be found 
with the approach
  of \citet{seljak2011}, in which the redshift-space density field
  is described in terms of the (density-weighted) velocity moments of
  the phase-space distribution function.  No trivial relation
  exists between our approach and such phase-space kinetic theory. 
  However, it is clear that in the definition of the distribution $\F$
  a role is played by the weighting over the number of pairs,
  suggesting that our description of $\P(\vpa)$ as a superposition of
  local distributions should rely ultimately on the phase-space
  dynamics of galaxy pairs. 

More in general, rather than focusing on just extracting the growth
rate of structure from RSD, we may want to consider the full
information contained in the velocity PDF, if we were able to recover it
from the data independently of the underlying
cosmology.  This would discriminate different gravity models even better.  
For example, it has been shown that the PDFs corresponding to the
$f(R)$ gravity models introduced by \citet{hu2007} are characterized
by a larger variance with respect to their $\Lambda$CDM counterparts
\citep[e.g.][]{fontanot2013}. 
Clearly, in a real galaxy survey we cannot directly measure the velocity distributions.
Still we could take advantage of our bivariate description to perform
a Monte Carlo sampling of the five parameters on which the bivariate
Gaussian depends. 
In essence, at each separation we are reducing the degrees of freedom
by compressing a continuous function into five numbers. 
The acceptance/rejection criterion of the random displacements in
parameter space can be obtained via the streaming-model by a standard
$\chi^2$ technique. 
In general, our five-parameter compression is not enough to
effectively measure the velocity PDFs on all scales (there are still
too many degrees of freedom). 
Nonetheless, by providing an appropriate functional form for
dependence of these parameters on separation, the degrees of freedom
can be further reduced. 
Also this possibility is currently being explored for a future paper.

We can then imagine further directions of development when considering
the streaming model expansion of Section~\ref{sec Streaming} . The
first important question to be answered is how large is the range of
convergence of the expansion of Eq. (\ref{eq
  expansion}).  
This could be both measured from simulations and discussed
theoretically, at least up to some given order and separation. 
If the expansion is indeed convergent over a significant range of
separations, more interesting questions arise.  First of all, 
    how many velocity moments do we need to recover the ``true''
    redshift-space correlation function on all scales? How many if we
    limit our analysis to some separation range, for example $s_\bot >
    10\hm1$Mpc? Can we predict them? 
    At least for the first two questions, a simple numerical approach
    can be easily applied. 

    Secondly, can we use this expansion to improve the
    description of how the BAO peak is distorted in redshift space? 
    This can be explored, for example, by substituting an ad hoc
    functional form for the baryonic peak into Eq. (\ref{eq
      expansion}), thus obtaining an analytic expression for the
    deviation from the linear Kaiser model, Eq. (\ref{eq Kaiser
      limit}), as a function of the velocity moments. 
    Given the precision of current and, even more, future BAO
    measurements from galaxy clustering, some insight into this issue
    might become crucial to avoid systematic effects on the peak
    position.    

    Thirdly, can we use the streaming-model expansion to directly measure the
    first velocity moments from the data?

Finally, we additionally note that the approach to RSD presented in this work and,
more in general, all those based on the streaming model, are well
suited to deal with the issue of velocity bias, since the contribution
of velocity is explicit. 
This is an important feature in the perspective of more and more
precise measurements that will require greater control over systematic
effects. 

\section*{Acknowledgments}

We thank J. Bel, S. de la Torre, V. Desjacques, J. Peacock,
R. Scoccimarro and R. Sheth
for useful discussions.
DB and MC acknowledge financial support by the University of
Milano through a PhD fellowship.  This work has been developed in the
framework of the ``Darklight'' program, supported by the European
Research Council through an Advanced Research Grant to LG
(Project \# 291521).  

\bibliographystyle{./mn2e}
\bibliography{./biblio_db}

\begin{thebibliography}{38}
\expandafter\ifx\csname natexlab\endcsname\relax\def\natexlab#1{#1}\fi

\bibitem[{{Bianchi} {et~al}\mbox{.}(2012){Bianchi}, {Guzzo}, {Branchini},
  {Majerotto}, {de la Torre}, {Marulli}, {Moscardini}, \&
  {Angulo}}]{bianchi2012}
{Bianchi} D., {Guzzo} L., {Branchini} E., {Majerotto} E., {de la Torre} S.,
  {Marulli} F., {Moscardini} L., {Angulo} R.~E., 2012, \mnras, 427, 2420

\bibitem[{{Davis} \& {Peebles}(1983)}]{davis1983}
{Davis} M., {Peebles} P.~J.~E., 1983, \apj, 267, 465

\bibitem[{{de la Torre} \& {Guzzo}(2012)}]{delatorre2012}
{de la Torre} S., {Guzzo} L., 2012, \mnras, 427, 327

\bibitem[{{Fisher}(1995)}]{fisher1995}
{Fisher} K.~B., 1995, \apj, 448, 494

\bibitem[{{Fisher} {et~al}\mbox{.}(1994){Fisher}, {Davis}, {Strauss}, {Yahil},
  \& {Huchra}}]{fisher1994b}
{Fisher} K.~B., {Davis} M., {Strauss} M.~A., {Yahil} A., {Huchra} J.~P., 1994,
  \mnras, 267, 927

\bibitem[{{Fontanot} {et~al}\mbox{.}(2013){Fontanot}, {Puchwein}, {Springel},
  \& {Bianchi}}]{fontanot2013}
{Fontanot} F., {Puchwein} E., {Springel} V., {Bianchi} D., 2013, \mnras, 436,
  2672

\bibitem[{{Guzzo} {et~al}\mbox{.}(2008){Guzzo}, {Pierleoni}, {Meneux},
  {Branchini}, {Le F{\`e}vre}, {Marinoni}, {Garilli}, {Blaizot}, {De Lucia},
  {Pollo}, {McCracken}, {Bottini}, {Le Brun}, {Maccagni}, {Picat},
  {Scaramella}, {Scodeggio}, {Tresse}, {Vettolani}, {Zanichelli}, {Adami},
  {Arnouts}, {Bardelli}, {Bolzonella}, {Bongiorno}, {Cappi}, {Charlot},
  {Ciliegi}, {Contini}, {Cucciati}, {de la Torre}, {Dolag}, {Foucaud},
  {Franzetti}, {Gavignaud}, {Ilbert}, {Iovino}, {Lamareille}, {Marano},
  {Mazure}, {Memeo}, {Merighi}, {Moscardini}, {Paltani}, {Pell{\`o}},
  {Perez-Montero}, {Pozzetti}, {Radovich}, {Vergani}, {Zamorani}, \&
  {Zucca}}]{guzzo2008}
{Guzzo} L. {et~al.}, 2008, \nat, 451, 541

\bibitem[{{Hamilton}(1992)}]{hamilton1992}
{Hamilton} A.~J.~S., 1992, \apjl, 385, L5

\bibitem[{{Hamilton}(1998)}]{hamilton1998}
{Hamilton} A.~J.~S., 1998, in Astrophysics and Space Science Library, Vol. 231,
  The Evolving Universe, {D.~Hamilton}, ed., pp. 185--+

\bibitem[{{Hawkins} {et~al}\mbox{.}(2003){Hawkins}, {Maddox}, {Cole}, {Lahav},
  {Madgwick}, {Norberg}, {Peacock}, {Baldry}, {Baugh}, {Bland-Hawthorn},
  {Bridges}, {Cannon}, {Colless}, {Collins}, {Couch}, {Dalton}, {De Propris},
  {Driver}, {Efstathiou}, {Ellis}, {Frenk}, {Glazebrook}, {Jackson}, {Jones},
  {Lewis}, {Lumsden}, {Percival}, {Peterson}, {Sutherland}, \&
  {Taylor}}]{hawkins2003}
{Hawkins} E. {et~al.}, 2003, \mnras, 346, 78

\bibitem[{{Hu} \& {Sawicki}(2007)}]{hu2007}
{Hu} W., {Sawicki} I., 2007, \prd, 76, 064004

\bibitem[{{Juszkiewicz}, {Fisher} \& {Szapudi}(1998){Juszkiewicz}, {Fisher}, \&
  {Szapudi}}]{juszkiewicz1998}
{Juszkiewicz} R., {Fisher} K.~B., {Szapudi} I., 1998, \apjl, 504, L1

\bibitem[{{Kaiser}(1984)}]{kaiser1984}
{Kaiser} N., 1984, \apjl, 284, L9

\bibitem[{{Kaiser}(1987)}]{kaiser1987}
{Kaiser} N., 1987, \mnras, 227, 1

\bibitem[{{Kwan}, {Lewis} \& {Linder}(2012){Kwan}, {Lewis}, \&
  {Linder}}]{kwan2012}
{Kwan} J., {Lewis} G.~F., {Linder} E.~V., 2012, \apj, 748, 78

\bibitem[{{Laureijs} {et~al}\mbox{.}(2011){Laureijs}, {Amiaux}, {Arduini},
  {Augu{\`e}res}, {Brinchmann}, {Cole}, {Cropper}, {Dabin}, {Duvet}, {Ealet},
  \& et~al.}]{laureijs2011}
{Laureijs} R. {et~al.}, 2011, ArXiv e-prints

\bibitem[{{Lewis}, {Challinor} \& {Lasenby}(2000){Lewis}, {Challinor}, \&
  {Lasenby}}]{lewis2000}
{Lewis} A., {Challinor} A., {Lasenby} A., 2000, \apj, 538, 473

\bibitem[{{Okumura} \& {Jing}(2011)}]{okumura2011}
{Okumura} T., {Jing} Y.~P., 2011, \apj, 726, 5

\bibitem[{{Peacock}(1999)}]{peacock1999}
{Peacock} J.~A., 1999, {Cosmological Physics}

\bibitem[{{Peacock} {et~al}\mbox{.}(2001){Peacock}, {Cole}, {Norberg}, {Baugh},
  {Bland-Hawthorn}, {Bridges}, {Cannon}, {Colless}, {Collins}, {Couch},
  {Dalton}, {Deeley}, {De Propris}, {Driver}, {Efstathiou}, {Ellis}, {Frenk},
  {Glazebrook}, {Jackson}, {Lahav}, {Lewis}, {Lumsden}, {Maddox}, {Percival},
  {Peterson}, {Price}, {Sutherland}, \& {Taylor}}]{peacock2001a}
{Peacock} J.~A. {et~al.}, 2001, \nat, 410, 169

\bibitem[{{Percival} \& {White}(2009)}]{percival2009}
{Percival} W.~J., {White} M., 2009, \mnras, 393, 297

\bibitem[{{Reid} {et~al}\mbox{.}(2012){Reid}, {Samushia}, {White}, {Percival},
  {Manera}, {Padmanabhan}, {Ross}, {S{\'a}nchez}, {Bailey}, {Bizyaev},
  {Bolton}, {Brewington}, {Brinkmann}, {Brownstein}, {Cuesta}, {Eisenstein},
  {Gunn}, {Honscheid}, {Malanushenko}, {Malanushenko}, {Maraston}, {McBride},
  {Muna}, {Nichol}, {Oravetz}, {Pan}, {de Putter}, {Roe}, {Ross}, {Schlegel},
  {Schneider}, {Seo}, {Shelden}, {Sheldon}, {Simmons}, {Skibba}, {Snedden},
  {Swanson}, {Thomas}, {Tinker}, {Tojeiro}, {Verde}, {Wake}, {Weaver},
  {Weinberg}, {Zehavi}, \& {Zhao}}]{reid2012}
{Reid} B.~A. {et~al.}, 2012, \mnras, 426, 2719

\bibitem[{{Reid} {et~al}\mbox{.}(2014){Reid}, {Seo}, {Leauthaud}, {Tinker}, \&
  {White}}]{reid2014}
{Reid} B.~A., {Seo} H.-J., {Leauthaud} A., {Tinker} J.~L., {White} M., 2014,
  \mnras, arXiv preprint, 1404.3742

\bibitem[{{Reid} \& {White}(2011)}]{reid2011}
{Reid} B.~A., {White} M., 2011, \mnras, 417, 1913

\bibitem[{{Riebe} {et~al}\mbox{.}(2013){Riebe}, {Partl}, {Enke},
  {Forero-Romero}, {Gottl{\"o}ber}, {Klypin}, {Lemson}, {Prada}, {Primack},
  {Steinmetz}, \& {Turchaninov}}]{riebe2013}
{Riebe} K. {et~al.}, 2013, Astronomische Nachrichten, 334, 691

\bibitem[{{Ross} {et~al}\mbox{.}(2007){Ross}, {da {\^A}ngela}, {Shanks},
  {Wake}, {Cannon}, {Edge}, {Nichol}, {Outram}, {Colless}, {Couch}, {Croom},
  {de Propris}, {Drinkwater}, {Eisenstein}, {Loveday}, {Pimbblet}, {Roseboom},
  {Schneider}, {Sharp}, \& {Weilbacher}}]{ross2007}
{Ross} N.~P. {et~al.}, 2007, \mnras, 381, 573

\bibitem[{{Samushia} {et~al}\mbox{.}(2014){Samushia}, {Reid}, {White},
  {Percival}, {Cuesta}, {Zhao}, {Ross}, {Manera}, {Aubourg}, {Beutler},
  {Brinkmann}, {Brownstein}, {Dawson}, {Eisenstein}, {Ho}, {Honscheid},
  {Maraston}, {Montesano}, {Nichol}, {Roe}, {Ross}, {S{\'a}nchez}, {Schlegel},
  {Schneider}, {Streblyanska}, {Thomas}, {Tinker}, {Wake}, {Weaver}, \&
  {Zehavi}}]{samushia2014}
{Samushia} L. {et~al.}, 2014, ArXiv e-prints, 439, 3504

\bibitem[{{Scoccimarro}(2004)}]{scoccimarro2004b}
{Scoccimarro} R., 2004, \prd, 70, 083007

\bibitem[{{Seljak} \& {McDonald}(2011)}]{seljak2011}
{Seljak} U., {McDonald} P., 2011, \jcap, 11, 39

\bibitem[{{Sheth}(1996)}]{sheth1996}
{Sheth} R.~K., 1996, \mnras, 279, 1310

\bibitem[{{Taruya}, {Nishimichi} \& {Saito}(2010){Taruya}, {Nishimichi}, \&
  {Saito}}]{taruya2010}
{Taruya} A., {Nishimichi} T., {Saito} S., 2010, \prd, 82, 063522

\bibitem[{{Tinker}(2007)}]{tinker2007}
{Tinker} J.~L., 2007, \mnras, 374, 477

\bibitem[{{Tinker}, {Weinberg} \& {Zheng}(2006){Tinker}, {Weinberg}, \&
  {Zheng}}]{tinker2006}
{Tinker} J.~L., {Weinberg} D.~H., {Zheng} Z., 2006, \mnras, 368, 85

\bibitem[{{Vallinotto} \& {Linder}(2013)}]{vallinotto2013}
{Vallinotto} A., {Linder} E.~V., 2013, ArXiv e-prints

\bibitem[{{White}, {Song} \& {Percival}(2009){White}, {Song}, \&
  {Percival}}]{white2009}
{White} M., {Song} Y., {Percival} W.~J., 2009, \mnras, 397, 1348

\bibitem[{{Zhang} {et~al}\mbox{.}(2007){Zhang}, {Liguori}, {Bean}, \&
  {Dodelson}}]{zhang2007}
{Zhang} P., {Liguori} M., {Bean} R., {Dodelson} S., 2007, PhysRevLett, 99

\bibitem[{{Zhang}, {Pan} \& {Zheng}(2013){Zhang}, {Pan}, \&
  {Zheng}}]{zhang2013}
{Zhang} P., {Pan} J., {Zheng} Y., 2013, \prd, 87, 063526

\bibitem[{{Zu} \& {Weinberg}(2013)}]{zu2013}
{Zu} Y., {Weinberg} D.~H., 2013, \mnras, 431, 3319

\end{thebibliography}

\appendix

\section{Derivation of the moments of the overall velocity distribution $\P$}\label{app moms}

We sketch here the derivation of the 3-rd moment of $\P$ as a function of the central moments of $P_L$ and $\F$.
We consider the most general case in which no assumptions are made on $\P_L$ and $\F$ (see Table \ref{tab moms}),
\begin{align}
\label{eq moms reminder}
 m^{(3)} = & {M_0}^3 + 6 M_1 C^{(2)}_{01} + 3 M_0 \left({M_1}^2 + C^{(2)}_{00} + C^{(2)}_{11}\right) \nonumber \\
 & + C^{(3)}_{000} + 3C^{(3)}_{011} + \langle c^{(3)}_L \rangle \ .  
\end{align}
All other moments can be obtained in a similar way. \\
Under the GG assumption, it is also possible to provide the moment generating function (which will be presented in a future work), so that the moments can be computed iteratively to any order. \\
In the following, we focus on the derivation of the most ``exotic'' terms of Eq. (\ref{eq moms reminder}), namely the correlation term $6M_1C^{(2)}_{01}$, the term contributed by the tensorial skewness $C^{(3)}_{000}+3C^{(3)}_{011}$ and the local-skewness term $\langle c^{(3)}_L \rangle$.
Under the GG assumption tensorial and local skewness are set to zero by definition and the only contribution to the skewness of $\P$ is given by the correlation term.
The key concept in the following calculations consists of completing squares and cubes.
We have
\begin{align}
 m^{(3)} = & \int dv \ v^3 \ \P(v) = \int dv \ v^3 \ \left\langle \P_L(v|\mu,\sigma) \right\rangle \nonumber \\
 = & \ \bigg\langle \int dv \ [{(v-\mu)}^3 - (-3v^2\mu+3v\mu^2-\mu^3)] \nonumber \\
 & \times \P_L(v|\mu,\sigma) \bigg\rangle \nonumber \\ 
 \label{eq summand1}
 = & \left\langle \int dv \ \P_L(v|\mu,\sigma) \ {(v-\mu)}^3 \right\rangle \\
 \label{eq summand2}
 & + \left\langle \int dv \ \P_L(v|\mu,\sigma) \ 3v^2\mu \right\rangle \\
 \label{eq summand3}
 & - \left\langle \int dv \ \P_L(v|\mu,\sigma) \ 3v\mu^2 \right\rangle \\
 \label{eq summand4}
 & + \left\langle \int dv \ \P_L(v|\mu,\sigma) \ \mu^3 \right\rangle \ .
\end{align}
Trivially, $\ \text{(\ref{eq summand1})} = \left\langle c^{(3)}_L \right\rangle$, $\ \text{(\ref{eq summand3})} = - 3 \left\langle \mu^3 \right\rangle$ and $\ \text{(\ref{eq summand4})} = \left\langle \mu^3 \right\rangle$.
This clarifies where the local-skewness term comes from.
As for the (\ref{eq summand2}) summand,
\begin{align}
 \text{(\ref{eq summand2})} = & \ 3\left\langle \mu \int dv \ \P_L(v|\mu,\sigma) \ v^2 \right\rangle \nonumber \\
 = & \ 3\left\langle \mu \int dv \ \P_L(v|\mu,\sigma) \ [{(v-\mu)}^2 - (-2v\mu+\mu^2)] \right\rangle \nonumber \\
 = & \ 3\left\langle \mu (\sigma^2 + 2\mu^2 - \mu^2) \right\rangle \nonumber \\
 = & \ 3\left\langle \mu\sigma^2\right\rangle + 3\left\langle \mu^3 \right\rangle \ .
\end{align}
Putting back together the summands, we get
\begin{align}
\label{eq calculations1}
 m^{(3)}  = \left\langle c^{(3)}_L \right\rangle + 3\left\langle \mu\sigma^2 \right\rangle + \left\langle \mu^3 \right\rangle \ .
\end{align}
To explicit the central (tensorial) moments of $\F$, we follow a similar procedure.
For example, the second summand can be written as
\begin{align}
\label{eq calculations2}
 3\left\langle \mu\sigma^2 \right\rangle = & \ 3\big\langle (\mu-M_0){(\sigma-M_1)}^2 \nonumber \\ 
 & - (-2\mu\sigma M_1 + \mu{M_1}^2 - \sigma^2 M_0 - 2\sigma M_0 M_1 - M_0{M_1}^2) \big\rangle \nonumber \\
  = & \ 3C^{(3)}_{011} + 6M_1\left\langle \mu\sigma \right\rangle - 3{M_1}^2\left\langle \mu \right\rangle + 3M_0\left\langle \sigma^2 \right\rangle \nonumber \\
 & + 6M_0 M_1\left\langle \sigma \right\rangle - 3M_0{M_1}^2 \ ,
\end{align}
which shows the origin of the $3C^{(3)}_{011}$ term.
The covariance term $6M_1C^{(2)}_{01}$ is then obtained by applying the same procedure to the second summand in the last row of Eq. (\ref{eq calculations2}), namely $6M_1\langle \mu\sigma \rangle$.
Similarly, from the third summand of Eq. (\ref{eq calculations1}), we recover $C^{(3)}_{000}$.\\
In general, when developing the right hand side of Eq. (\ref{eq calculations1}) polynomials in $M_k$ and $C^{(2)}_{kk}$ are produced: putting back together all the pieces, we eventually recover Eq. (\ref{eq moms reminder}).

\end{document}